\documentclass[12pt]{article}
\usepackage{hyperref}
\usepackage{amsmath,amssymb}
\usepackage{graphicx}
\usepackage{color}
\definecolor{myblue}{rgb}{0.14,0.11,0.49}
\definecolor{myred}{rgb}{0.74,0.22,0.15}
\definecolor{mygreen}{rgb}{0.05,0.52,0.42}
\definecolor{myyellow}{rgb}{0.96,0.92,0.13}
\definecolor{myorange}{rgb}{1,0.61,0.36}
\definecolor{mypurple}{rgb}{0.71,0.02,1}
\definecolor{noir}{gray}{0.} % black

\newcommand{\Couleur}[1]{\textcolor{noir}{#1}}

\definecolor{htc}{rgb}{1,1,1} % heading text colour
\newcommand{\Mat}[1]{{{\boldsymbol{#1}}}}
\newcommand{\abs}[1]{\left\vert#1\right\vert}
\def\be{\begin{equation}}
\def\ee{\end{equation}}
\def\bea{\begin{eqnarray}}
\def\eea{\end{eqnarray}}
\def\bc{\begin{center}}
\def\ec{\end{center}}
\def\bi{\begin{itemize}}
\def\ei{\end{itemize}}
\def\bs{\begin{slide}}
\def\es{\end{slide}}
\def\dd{\mathrm{d}}
\def\iC{\mathrm{i}}
\def\noi{\noindent}

\title{Spectral energy density in an axisymmetric galaxy as predicted by an analytical model for the Maxwell field}

\author{
Mayeul Arminjon\\
\small\it Univ. Grenoble Alpes, CNRS, Grenoble INP%\footnote{\ Institute of Engineering Univ. Grenoble Alpes}
, 3SR, F-38000 Grenoble, France\\
\small\it  E-mail: Mayeul.Arminjon@3sr-grenoble.fr.
\small\it Orcid: 0000-0002-7035-351X
} 
\date{}
\begin{document}
%%%%%%%%%%%%%%%%%%%%%%%%%%%%%%%%%%%%%%%%%%%%%%%%%%%%%%%%%%%%%%%%%%%%%%%%%%%%%%%%

\maketitle

\noi Short title: {\it Spectral energy density in a galaxy from an exact Maxwell field}
 
\begin{abstract}

\noi An analytical model for the Maxwell radiation field in an axisymmetric galaxy, proposed previously, is first checked for its predictions of the spatial variation of the spectral energy distributions (SEDs) in our Galaxy. First, the model is summarized. It is now shown how to compute the SED with this model. Then the model is adjusted by asking that the SED predicted at our local position in the Galaxy coincide with the available observations. Finally the first predictions of the model for the spatial variation of the SED in the Galaxy are compared with those of a radiation transfer model. We find that the two predictions do not differ too much. This indicates that, in a future work, it should be possible with the present model to check if the ``interaction energy" predicted by an alternative, scalar theory of gravitation, contributes to the dark matter.\\ 

\noi {\bf Keywords:} Disk galaxy; interstellar radiation field; axial symmetry; electromagnetic field; Milky Way.\\

\end{abstract}

			     % (``~`` are extra hard spaces)

%%%%%%%%%%%%%%%%%%%%%%%%%%%%%%%%%%%%%%%%%%%%%%%%%%%%%%%%%%%%%%%%%%%%%%%%%%%%%%%%
\section{Introduction}
%%%%%%%%%%%%%%%%%%%%%%%%%%%%%%%%%%%%%%%%%%%%%%%%%%%%%%%%%%%%%%%%%%%%%%%%%%%%%%%%

The interstellar radiation field (ISRF) is a very important physical characteristic of a galaxy, and there is indeed a rich astrophysical literature that studies the modeling of the ISRF; see e.g. \cite{Gordon-et-al2001, PorterStrong2005, Popescu-et-al2011, Robitaille2011, Maciel2013, Popescu-et-al2017} and references therein. This subject is of importance not only in itself, but also because the radiation field produced by the stars and other bright objects has a strong interaction with the cosmic rays, which are another important component of the interstellar medium. The necessity of a modeling is obvious, since we can directly measure the radiation field only in the solar system, in fact in the neighborhood of the Earth \cite{Henry-et-al1980,Arendt-et-al1998,Finkbeiner-et-al1999}. The ISRF is significantly affected by its interaction with interstellar dust (e.g. \cite{KylafisBahcall1987, Witt-et-al1992}). In order to account for this interaction, equations for radiative transfer are solved: either by using Monte-Carlo methods, that simulate in a probabilistic framework the propagation of photons, and their absorption and scattering by dust, in a discretized spatial domain (e.g. \cite{Gordon-et-al2001, Robitaille2011}); or, by using ``ray-tracing" algorithms that calculate the variation, due to absorption and scattering by dust, of the radiation's specific intensity along a finite set of directions (e.g. \cite{Popescu-et-al2011, Robitaille2011, Popescu-et-al2017}). Thus, until now, the electromagnetic (EM) field, made of coupled electric and magnetic fields, has not been considered in the studies of the ISRF in galaxies, even less an EM field that would be a solution of the Maxwell equations --- as should actually be the case for the real ISRF, however. \\

Obtaining for the ISRF such an exact solution of the Maxwell equations is somewhat demanding from both the mathematical and the numerical points of view, and has been the main purpose of a recent work \cite{A61}. A decisive motivation for that recent work was to prepare for checking an exotic prediction of an alternative, scalar theory of gravitation: namely, that the total energy(-momentum-stress) tensor, which provides the source of the gravitational field, is not in general the sum of the energy tensor of the EM field and the energy tensor of the material medium producing it. Instead, each time that an EM field is present together with a variable gravitational field (which is the general case), there should appear an additional energy tensor: the ``interaction" energy tensor, $\Mat{T}_\mathrm{inter}$ \cite{A57}. This speculative energy tensor $\Mat{T}_\mathrm{inter}$, being thus present virtually everywhere according to that theory, and having clearly an exotic character, could contribute to the dark matter. In order to calculate $\Mat{T}_\mathrm{inter}$, and to see if its distribution could be similar to dark matter halos, it is necessary to be able to calculate the EM field and its first order derivatives \cite{A57}, whence the work \cite{A61}. In the latter work, a model has been formulated. It is based on an analytical representation of the general solution of the source-free Maxwell equations in the axisymmetric case \cite{A60}. Moreover, the model has been numerically implemented and has passed a numerical validation test based on an exact solution \cite{A61}.\\

The aim of the present paper is to make a first check of the predictions of that ``Maxwell model of the interstellar radiation field" for the spectral energy distribution (SED) and its spatial variation in a galaxy. Such SEDs are a very important output of the mainstream models for the analysis of the ISRF and, as was recalled at the beginning of this paper, those models depend essentially of the radiation transfer (mainly through dust). Note that, in contrast, the present model \cite{A61} does not directly take into account any radiation transfer. Section \ref{Model} summarizes that model. Section \ref{SED-Calculn} details the calculation of the SED in that model. Section \ref{Results} presents the results obtained with that model for a simulation of the Milky Way. To obtain these results, the model is adjusted on the locally observed SED. In this way, it does indirectly take into account the radiation transfer, because the observed SED is indeed affected by the latter.

\section{The Maxwell model of the interstellar radiation field} \label{Model}

In this section, we give a new presentation of the model which was built in Ref. \cite{A61}, and that uses results of Ref. \cite{A60} and references therein. This presentation uses an easy extension of the result \cite{A60}. Such an extension was used implicitly in Ref. \cite{A61}, and is explicitly formulated in Appendix \ref{Extension}.

\subsection{General assumptions for the Maxwell field}

We make two assumptions regarding the form of the Maxwell radiation field $({\bf E},{\bf B})$ (with ${\bf E}$ the electric field and ${\bf B}$ the magnetic field):\\

i) We assume that it is an {\it axially symmetric} field. This is a relevant first approximation for many galaxies, even though it neglects some aspects like the arm structure of a spiral galaxy. Axial symmetry is often assumed in the literature on modeling the ISRF in galaxies and/or its interaction with dust (e.g. \cite{PorterStrong2005,Popescu-et-al2011, Popescu-et-al2017, KylafisBahcall1987}). \\

ii) We search for a {\it source-free} solution of the Maxwell equations. This is because we seek to describe the EM field at the galactic scale, not the local EM field in the neighborhood of the stars that are the primary sources of that field.

\subsection{The explicit representation for axisymmetric free Maxwell fields}\label{Explicit}

This is based on the following recent \hypertarget{Theorem}{{\it Theorem} }\cite{A60}: any time-harmonic axisymmetric solution $({\bf E},{\bf B})$ of the source-free Maxwell equations has a unique decomposition 
\be\label{Decompos}
{\bf E} = {\bf E}_1 + {\bf E}'_2,\qquad {\bf B} = {\bf B}_1 + {\bf B}'_2,
\ee
where $({\bf E}_1, {\bf B}_1)$ has the form (in cylindrical coordinates $\rho ,\phi ,z$ corresponding with the symmetry axis $Oz$):
\bea\label{Bfi}
\Couleur{B_\phi} & = & \Couleur{-\frac{\partial \Psi}{\partial \rho }},\qquad \Couleur{E_\phi =0},\\
\nonumber\\
\label{Erho}
\Couleur{E_\rho} & = & \Couleur{\iC \frac{c^2}{\omega } \frac{\partial^2 \Psi}{\partial \rho \,\partial z}},\qquad \Couleur{B_\rho =0},\\
\nonumber\\
\label{Ez}
\Couleur{E_z} & = & \Couleur{\iC \frac{c^2}{\omega } \frac{\partial^2 \Psi}{\partial z^2 } + \iC \omega \Psi}, \qquad \Couleur{B_z=0};
\eea
and where $({\bf E}'_2, {\bf B}'_2)$ is deduced from a solution $({\bf E}_2, {\bf B}_2)$ having just that same form (\ref{Bfi})--(\ref{Ez}), by the EM duality --- i.e.,
\be\label{dual}
{\bf E}' = c{\bf B}, \quad {\bf B}' = -{\bf E}/c.
\ee
(Here $c$ is the velocity of light.) Note that the dual solution (\ref{dual}) has the complementary zero and non-zero components as compared with the direct solution (\ref{Bfi})--(\ref{Ez}), i.e., $B'_\phi =E'_\rho =E'_z=0$, while $E'_\phi, B'_\rho,B'_z \ne 0$ in general. Both solutions $({\bf E}_1, {\bf B}_1)$ and $({\bf E}'_2, {\bf B}'_2)$ are time-harmonic with the same angular frequency $\omega $ as has $({\bf E}, {\bf B})$. In Eqs. (\ref{Bfi})--(\ref{Ez}), the potential $\Psi(t,\rho ,z)=e^{-\iC \omega t} \hat{\Psi } (\rho ,z)$ is a time-harmonic axisymmetric solution of the standard scalar wave equation, which also has the same frequency $\omega $ as has $({\bf E}, {\bf B})$; in general, the two solutions $({\bf E}_1, {\bf B}_1)$ and $({\bf E}_2, {\bf B}_2)$ correspond with two different scalar potentials $\Psi$ and $\Psi'$ \cite{A60}. The solution (\ref{Bfi})--(\ref{Ez}) actually derives from a vector potential having just its axial component $A_z=\Psi$ non-zero, i.e., ${\bf A} = \Psi {\bf e}_z$ \cite{A60,GAZR2014}. The Theorem \cite{A60} states the existence of the two potentials $\Psi$ and $\Psi'$, not their uniqueness. In Appendix \ref{Extension}, we extend this result to EM fields having a finite spectrum, which is easy, and allows us to give a more rigorous presentation of the model in Subsect. \ref{ModelStatement}.\\

If a time-harmonic, axisymmetric solution of the scalar wave equation %($\Psi \in\mathcal{S}^\omega$ in the notation of Appendix \ref{Extension})
 is {\it totally propagating,} it has the following explicit form \cite{GAZR2014,ZR_et_al2008}:
\be\label{psi_monochrom}
\Couleur{\psi _{\omega\ S} \,(t,\rho,z) = e^{-\iC \omega t} \int _{-K} ^{+K}\ J_0\left(\rho \sqrt{K^2-k^2}\right )\ e^{\iC k \, z} \,S(k)\, \dd k},
\ee
with $\omega $ the angular frequency, $K:=\omega /c$, $J_0$ the first-kind Bessel function of order $0$, and where $S$ is some (generally complex) function of $k \in [-K,+K]$. ``Totally propagating" means, in short, that $\Psi$ has no ``evanescent" component \cite{GAZR2014}; a detailed explanation is given in Ref. \cite{A61}. In turn, we shall define simply that a time-harmonic, axisymmetric EM field $\Mat{F}=({\bf E},{\bf B})$ is totally propagating if the two potentials $\Psi $ and $\Psi'$, whose existence is guaranteed by the \hyperlink{Theorem}{Theorem} recalled above, can themselves be chosen totally propagating, i.e., having the form (\ref{psi_monochrom}). 
\\

\subsection{Statement of the model}\label{ModelStatement}

We model a disk galaxy as a disk-like set of point-like ``stars".
\footnote{\ 
The corresponding points are got by pseudo-random generation of their cylindrical coordinates. Here, the same set of $16\times 16 \times 36$ triplets $(\rho ,z,\phi )$ was used as in Ref. \cite{A61}, thus $9216$ points. The distribution of $\rho $ and $z$ is approximately that valid for the star distribution in the Milky Way, and the distribution of $\phi $ is uniform, thus ensuring approximate axial symmetry.
}
Each of them is assumed to produce an EM field with a finite spectrum: that field derives, in the way defined in the previous section, from potentials that obey the scalar wave equation, more precisely ones with spherical symmetry:
\be\label{psi_spher'}
\psi _{\omega_j} \ (t,{\bf x}) = \frac{e^{\iC (K_j r-\omega_j t)}} {K_j r}, \qquad K_j:=\frac{\omega_j }{c}, \qquad r:=\abs{{\bf x}}. 
\ee
Thus, the sum of the potentials emitted by the different stars at the different frequencies is:
\be\label{sum_psi_spher_spectre}
\Psi _{({\bf x}_i)\ (\omega _j)\ (w_j)}\,(t,{\bf x}) = \sum _{i=1} ^{i_\mathrm{max}} \sum _{j=1} ^{N_\omega} \,w_j\,\frac{e^{\iC (K_j\, r_i-\omega_j t)}} {K_j \ r_i}.
\ee
Here $\,r_i:=\abs{{\bf x}-{\bf x}_i}$, with $\,{\bf x}\,$ the spatial position at which the function is evaluated and $\,{\bf x}_i$ the spatial position of the $i$-th star; the numbers $\,w_j>0$, with $\,\sum_j w_j=1$, are the weights affected to the different frequencies. 
\footnote{\ 
In Ref. \cite{A61}, these weights were noted $S'_j$, which now would clash with the functions $S'_j$ mentioned at the end of the present paper.
}  
Since the disk-like distribution of the ``stars" is built (approximately) axisymmetric \cite{A61}, the function (\ref{sum_psi_spher_spectre}) is also (approximately) axisymmetric. Moreover, this is a solution of the scalar wave equation, and it is clearly a totally propagating one, as is each of the individual spherical scalar waves in (\ref{sum_psi_spher_spectre}). It should thus be possible to put the function (\ref{sum_psi_spher_spectre}) --- at least approximately --- in the form of a sum, over the frequencies, of functions having the form (\ref{psi_monochrom}). This is done by the least-squares method applied on a finite set of events $\,(t,{\bf x})$ (with ${\bf x}={\bf x}(\rho ,\phi =0,z)$ restricted to the plane $\phi =0$, due to the axial symmetry), making a regular spatio-temporal grid $\{(t_l,\rho _m,z_p)\}$. We thus want to determine complex functions $S_j=S_j(k)\ (-K_j \le k \le +K_j)$, such that 
\be\label{Psi-simeq-Psi'}
\Psi _{({\bf x}_i)\ (\omega _j)\ (w_j)}\,(t,\rho ,z) \cong  \sum _{j} \,\psi _{\omega_j\ S_j}\,(t,\rho , z):=\Psi _{ (\omega _j)\ (S_j)}\,(t,\rho ,z),
\ee
where the sign $\cong$ indicates that the equality is in the sense of the least squares (on the relevant spatio-temporal grid, just mentioned). To the unique function on the r.h.s. of (\ref{Psi-simeq-Psi'}): $\Psi =\sum _{j} \,\psi _{\omega_j\ S_j}$, we may then associate: 
\bi
\item  either the EM field obtained by summing the time-harmonic contributions given by Eqs. (\ref{Bfi})--(\ref{Ez}), with $A_z:=\psi _{\omega_j\ S_j}$ --- that is, 
\be\label{F=Z_1(Psi)}
\Mat{F}_1 := Z_1(\Psi )
\ee 
(cf. Eq. (\ref{F_1,F_2}) in the Appendix);
\item or, the dual (\ref{dual}) of that field --- that is, 
\be\label{F=Z_2(Psi)}
\Mat{F}_2 := Z_2(\Psi );
\ee
\item or still, the sum of these two fields, that is 
\be\label{F_1+F_2_Psi'=Psi}
\Mat{F}= \Mat{F}_1+\Mat{F}_2 = Z_1(\Psi )+Z_2(\Psi ) :=Z(\Psi ,\Psi )
\ee
(cf. Eq. (\ref{F_1+F_2}) in the Appendix).
\ei

\subsection{Discretization}

To calculate the integrals in Eq. (\ref{psi_monochrom}) and in Eqs. (\ref{Bfi})--(\ref{Ez}) applied with functions $\psi _{\omega_j\ S_j}$ of the form (\ref{psi_monochrom}), we discretize the integration intervals $[-K_j,+K_j]$ and the functions $S_j$, and we use the ``Simpson $\frac{3}{8}$ composite rule" \cite{A61}. We set $K_0:=\omega _0/c$, with $\omega _0$ some (arbitrary) reference frequence, and
\be\label{k_n}
k_n := -K_0 + n\delta _0 \quad (n=0,...,N),\qquad \delta _0 := 2K_0/N,
\ee
\be\label{S_nj}
S_{n j} := S_j\left (\frac{\omega _j}{\omega _0}k_n\right )\quad (n=0,...,N,\quad j=1,..., N_\omega), 
\ee
\bea\label{a_j}
a_{n} & := & (3/8)\,\delta _0 \quad (n = 0\ \mathrm{or}\ n = N),\\
a_{n} & := & 2\times (3/8)\,\delta _0 \quad (\mathrm{mod}(n,3)=0 \ \mathrm{and}\ n \ne 0\ \mathrm{and}\ n \ne N),\\
a_{n} & := &  3\times (3/8)\,\delta _0\quad \mathrm{otherwise}.
\eea
With this discretization, the unknown functions $S_j\ (j=1,...,N_\omega )$ in Eq. (\ref{Psi-simeq-Psi'}) are characterized by the complex numbers $S_{n j}\ (n=0,...,N;\, j=1,..., N_\omega )$.  Thus, the equality in the sense of the least squares, schematized by Eq. (\ref{Psi-simeq-Psi'}), actually becomes \cite{A61}
\be\label{Psi'}
\Psi _{({\bf x}_i)\ (\omega _j)\ (w_j)}\,(t,\rho ,z) \cong \sum _{n=0} ^N \sum _{j = 1} ^{N_\omega } f_{n j}(t,\rho ,z) \,S_{n j},% + O\left(\frac{1}{N^4}\right),
\ee
with 
\be\label{f_nj}
f_{n j}(t,\rho ,z) = \frac{\omega _j}{\omega _0} a_n \,J_0\left(\rho \frac{\omega _j}{\omega _0} \sqrt{K_0^2 - k_n^2} \right) \exp \left[ \iC \left( \frac{\omega _j}{\omega _0} k_n z -\omega _j\, t\right ) \right ].
\ee 
The $S_{n j}$ 's are the solved-for parameters in the least-squares problem (\ref{Psi'}). \\

This gives us for the EM field (\ref{F=Z_1(Psi)})  \cite{A61}, as also for the field (\ref{F_1+F_2_Psi'=Psi}):
\be\label{Bphi'}
B_\phi (t,\rho ,z) = \sum _{n=0} ^N \sum _{j = 1} ^{N_\omega } R_n \,J_1\left(\rho \frac{\omega _j}{\omega _0} R_n \right) \,{\mathcal Re} \left[ F_{n j}(t,z)\right ] + O\left(\frac{1}{N^4}\right),
\ee
\be\label{Erho'}
E_\rho (t,\rho ,z) = \sum _{n=0} ^N \sum _{j = 1} ^{N_\omega } \frac{c^2}{\omega_0 } k_n R_n\,J_1\left(\rho \frac{\omega _j}{\omega _0} R_n \right) \,{\mathcal Re} \left[F_{n j}(t,z)\right ] + O\left(\frac{1}{N^4}\right) ,
\ee
\be\label{Ez'}
E_z (t,\rho ,z) = \sum _{n=0} ^N \sum _{j = 1} ^{N_\omega } \left(\frac{c^2}{\omega_0 } k_n^2 -\omega _0 \right ) \,J_0\left(\rho \frac{\omega _j}{\omega _0} R_n \right) \,{\mathcal Im} \left[F_{n j}(t,z) \right] + O\left(\frac{1}{N^4}\right),
\ee
with \ $R_n = \sqrt{K_0^2 - k_n^2}$ \ and
\be\label{F_nj}
F_{n j}(t,z) = \left(\frac{\omega _j}{\omega _0}\right)^2\, a_n \exp \left[ \iC \left( \frac{\omega _j}{\omega _0} k_n z -\omega _j\, t\right ) \right ]\,S_{n j}.
\ee
Even though these formulae follow from a discretization and from using an integration rule that is only approximate for the starting integral, they provide (without the $O\left(\frac{1}{N^4}\right)$ remainder) an exact solution of the source-free Maxwell equations --- see \S 3.3 in Ref. \cite{A61}. \\

The arguments of the Bessel functions $J_0$ and $J_1$ in Eqs. (\ref{f_nj}), (\ref{Bphi'})--(\ref{Ez'}), as well as the spatial part of the argument of the complex exponential in Eqs. (\ref{f_nj}) and (\ref{F_nj}), are of the order of $\rho /\lambda $ or $z/\lambda $. For a galaxy, these are numbers of the order of $10^{25}$, therefore the  numerical implementation of the model needs to use a precision better than quadruple precision --- which leads to long computation times \cite{A61}.

\section{Calculation of the spectral energy density}\label{SED-Calculn}

The volumic energy density of an EM field, in SI units, is
\be\label{Def-U}
U = \frac{\delta W}{\delta V} = \frac{\epsilon _0}{2}  \left({\bf E}^2 + c^2 {\bf B}^2 \right ),
\ee
with $\epsilon _0 = 1/(4\pi \times 9\times 10^9)$. In the relevant case with a frequency spectrum, the fields corresponding to the different frequencies do add, hence the quadratic expression (\ref{Def-U}) is not additive. The {\it time average} of such a quadratic expression is yet known to be additive in the case of a Fourier expansion \cite{L&L-118}. Let us check if this remains true in the present case of an arbitrary finite frequency spectrum $(\omega _j) \ (j=1,...,N_\omega)$. Let $F(t)$ be some component of the EM field at some given spatial position ${\bf x}$. We have, with some complex coefficients $C_j = C_j({\bf x})$:
\be\label{F(t)}
F(t) = {\mathcal Re} \left ( \sum _{j=1} ^{N_\omega } C_j e^{-\iC \omega _j t} \right ),
\ee
whence
\be\label{4F^2}
4F^2 = \left ( \sum _{j=1} ^{N_\omega } C_j e^{-\iC \omega _j t} + C_j^\ast  e^{\iC \omega _j t} \right )\left ( \sum _{k=1} ^{N_\omega } C_k e^{-\iC \omega _k t} + C_k^\ast  e^{\iC \omega _k t} \right ).
\ee
Unless $\omega _j+\omega _k=0$, the time average $\overline{e^{-\iC \left (\omega _j+\omega _k \right )t}}$ is zero. Unless $\omega _j-\omega _k=0$, the time average $\overline{e^{-\iC \left (\omega _j-\omega _k \right )t}}$ is zero. But $\omega _j+\omega _k=0$ does not occur, because $\omega _j>0$, and $\omega _j-\omega _k=0$ occurs only when $j=k$. Hence, (\ref{4F^2}) gives indeed the additivity for the time average of the square of a component admitting an expansion (\ref{F(t)}): 
\be\label{F^2}
\overline{F^2} = \frac{1}{4}\sum _{j=1} ^{N_\omega } C_j C_j^\ast + C_j^\ast C_j = \frac{1}{2}\sum _{j=1} ^{N_\omega } \abs{C_j} ^2.
\ee
For the EM field (\ref{F=Z_1(Psi)}), its components $B_\phi ,\, E_\rho ,\, E_z$ admit expansions (\ref{F(t)}) with coefficients $C^{(1)}_j$, $C^{(2)}_j$, $C^{(3)}_j$, respectively, which are directly read on Eqs. (\ref{Bphi'})--(\ref{Ez'}):
\be\label{C1}
C^{(1)}_j(\rho ,z) = \sum _{n=0} ^N R_n \,J_1\left(\rho \frac{\omega _j}{\omega _0} R_n \right) G_{nj},
\ee
\be\label{C2}
C^{(2)}_j (\rho ,z)= \sum _{n=0} ^N \frac{c^2}{\omega_0 } k_n R_n \,J_1\left(\rho \frac{\omega _j}{\omega _0} R_n \right) G_{nj},
\ee
\be\label{C3}
C^{(3)}_j (\rho ,z) = \sum _{n=0} ^N \left(\frac{c^2}{\omega_0 } k_n^2 -\omega _0 \right ) \,J_0\left(\rho \frac{\omega _j}{\omega _0} R_n \right) \left ( -\iC G_{nj}\right ),
\ee
with \ $R_n = \sqrt{K_0^2 - k_n^2}$ \ and
\be\label{G_nj}
G_{n j}(z) = \exp \left( \iC \omega _j\, t \right )  F_{n j}(t,z) = \left(\frac{\omega _j}{\omega _0}\right)^2\, a_n \exp \left ( \iC \frac{\omega _j}{\omega _0} k_n z  \right )\,S_{n j}.
\ee
From  (\ref{Def-U}) and (\ref{F^2}), we thus get for the field (\ref{F=Z_1(Psi)}) a (time-averaged) energy density with a discrete spectrum:
\be\label{Udiscrete}
\overline{U}(\rho ,z) :=\overline{\frac{\delta W}{\delta V}} = \sum _{j=1} ^{N_\omega } u_j, \qquad u_j:= \frac{1}{4} \sum _{q=1} ^3 \alpha _q \abs{C^{(q)}_j}^2,
\ee
where the coefficients $C^{(q)}_j\ (q=1,2,3)$  are given by Eqs. (\ref{C1})--(\ref{C3}), $\alpha _1=\epsilon _0 c^2$, $\alpha _2=\alpha _3=\epsilon _0$. %(We will henceforth omit the bar that indicates the time average...)
As to the field (\ref{F=Z_2(Psi)}), it is the dual (\ref{dual}) of the field (\ref{F=Z_1(Psi)}). Therefore, the energy density (\ref{Def-U}) for the field (\ref{F=Z_2(Psi)}) is given by the same Eq.  (\ref{Udiscrete}). Finally, the energy density for the field (\ref{F_1+F_2_Psi'=Psi}) is hence just the double of this. Now, in the next section, the model will be adjusted so as to exactly describe the measured local EM spectrum, i.e., the values $u_j({\bf x}_\mathrm{loc})$ will be imposed to be the ones measured at our local position ${\bf x}_\mathrm{loc}$ in the Galaxy. In this context, the precise choice of the model, i.e., Eq. (\ref{F=Z_1(Psi)}), or (\ref{F=Z_2(Psi)}), or (\ref{F_1+F_2_Psi'=Psi}), is therefore totally neutral.

%%%%%%%%%%%%%%%%%%%%%%%%%%%%%%%%%%%%%%%%%%%%%%%%%%%%%%%%%%%%%%%%%%%%%%%%%%%%%%%%
\section{Results and comparison with a radiation transfer model} \label{Results}
%%%%%%%%%%%%%%%%%%%%%%%%%%%%%%%%%%%%%%%%%%%%%%%%%%%%%%%%%%%%%%%%%%%%%%%%%%%%%%%%

\begin{figure}[ht]
\centerline{\includegraphics[height=10cm]{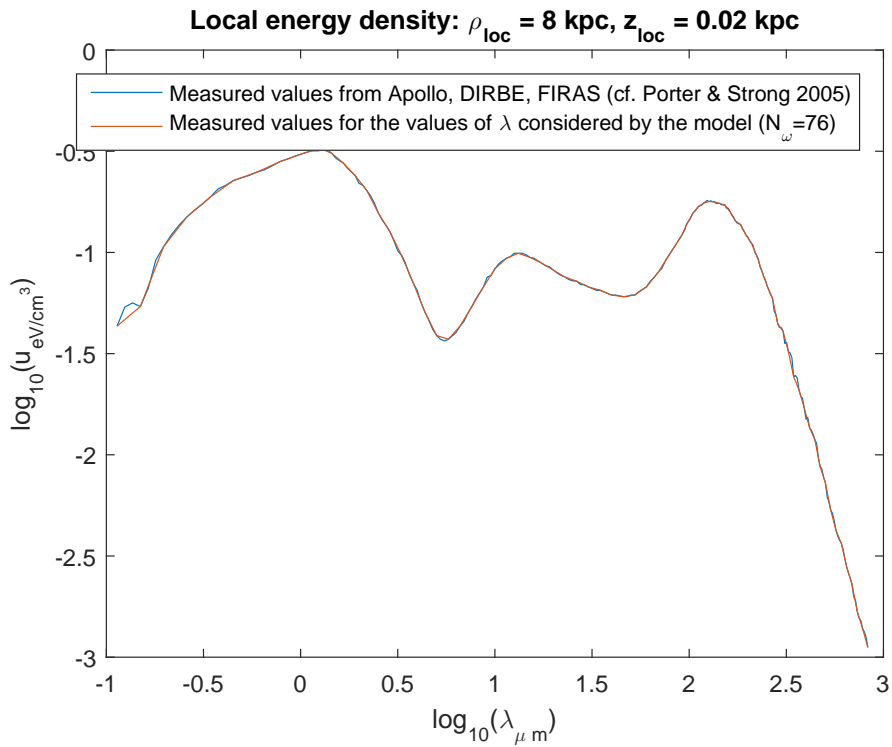}}
\caption{SED as measured by space missions. Between the data of Apollo (ending at $\mathrm{log}_{10}\lambda \simeq -0.8$) %-0.85$ 
and DIRBE (beginning at $\mathrm{log}_{10}\lambda \simeq +0.1$)%+0.097$)
, the curve has been interpolated.}
\label{Local}
\end{figure}

We first explain the final adjustment of the model. The scalar wave $\Psi _{({\bf x}_i)\ (\omega _j)\ (w_j)}$ given by Eq. (\ref{sum_psi_spher_spectre}) depends on the numbers $w_j>0$, with $\sum_j w_j=1$. Each of them represents the relative weight of the corresponding frequency $\omega _j$ --- though in some average sense, since $w_j$ is common to all ``stars" at the points ${\bf x}_i$. There is no exact way to assign values to these relative weights $w_j$, which, together with the set $({\bf x}_i)$ of the positions of the stars, determine the functions $S_j$ on the r.h.s. of Eq. (\ref{Psi-simeq-Psi'}) --- or more exactly, determine the discretized values $S_{n j}$ in Eqs. (\ref{S_nj}) and (\ref{Psi'}). In turn, the values $S_{n j}\ (n=0,...,N)$, for a given value of the frequency index $j$, determine the intensity of the $\omega _j$-component of the scalar wave obtained after the fitting (the r.h.s. of (\ref{Psi'})), and thus determine the intensity of the $\omega _j$-component of the EM field (\ref{Bphi'})--(\ref{Ez'}). However, since we start only from relative weights $w_j$, this determination can be only up to a multiplying factor, and since the very values of the $w_j$ 's are assigned somewhat arbitrarily, that determination of the $S_{n j}$ 's $\ (n=0,...,N)$ is really up to a factor depending on $j$, $\xi _j>0$.
\footnote{\
This would be even clearer if we would separate the least-squares adjustment (\ref{Psi'}) into the different frequencies $\omega _j$, which is actually easy and even would allow us to eliminate the common time dependence $\exp(- \iC \omega _j t)$ on both sides, thus eliminating the time variable in the adjustment. This different procedure will be numerically tested in a future work.
}
 We determine the values $\xi_j$ by imposing that the values $u_j$ given by Eq. (\ref{Udiscrete}), calculated at our local position ${\bf x}_\mathrm{loc}$ in the Galaxy, be equal to the ones measured. We take $\rho_\mathrm{loc} = 8$ kpc and $z_\mathrm{loc} = 0.02$ kpc (e.g. \cite{Majaess-et-al2009}). We took the local values of the SED from Fig. 1 of Ref. \cite{PorterStrong2005}, using the software CurveUnscan to digitalize the curves. Those local values were determined from the Apollo 17 mission \cite{Henry-et-al1980} and from the COBE mission: the DIRBE experiment \cite{Arendt-et-al1998} and the FIRAS experiment \cite{Finkbeiner-et-al1999}. Figure \ref{Local} shows the corresponding SED. For the calculations, in order to save computer time, we extracted 76 values of $\lambda $ from the complete list of values of $\lambda $ in the curve; the corresponding points %$(\log\,\lambda ,\lambda u(\lambda ))$ 
are also shown on Fig. \ref{Local}.
%(Here $u(\lambda )$ is the continuous SED, such that the (time-averaged) volumic energy in a band $[\lambda _1,\lambda _2]$ is $\Delta \overline{U} =\int_{\lambda _1} ^{\lambda _2} u(\lambda ) \dd \lambda $. Note that it is indeed $\lambda u(\lambda )$ which has the dimension of a volumic energy density, as have the $u_j$ 's in Eq. (\ref{Udiscrete})$_1$.) %\footnote{\If $u(\lambda )$ is the SED, such that the (time-averaged) volumic energy in a band $[\lambda _1,\lambda _2]$ is $\Delta \overline{U} =\int_{\lambda _1} ^{\lambda _2} u(\lambda ) \dd \lambda $, we have als o $\Delta \overline{U} = \int_{\log \lambda _1} ^{\log \lambda _2} \lambda u(\lambda ) \dd (\log \lambda)$ }
\\

\begin{figure}[ht]
\centerline{\includegraphics[height=10cm]{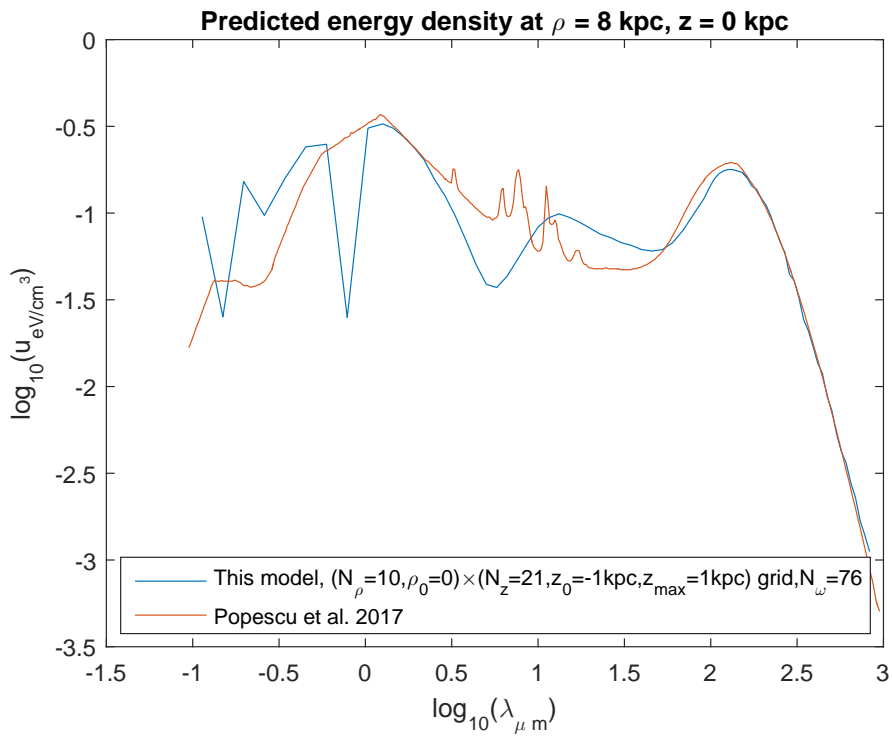}}
\caption{SED at $\rho =8\,$kpc, $z=0\,$kpc, according to this model or a radiation transfer model.}
\label{SED_8_0}
\end{figure}

\begin{figure}[ht]
\centerline{\includegraphics[height=10cm]{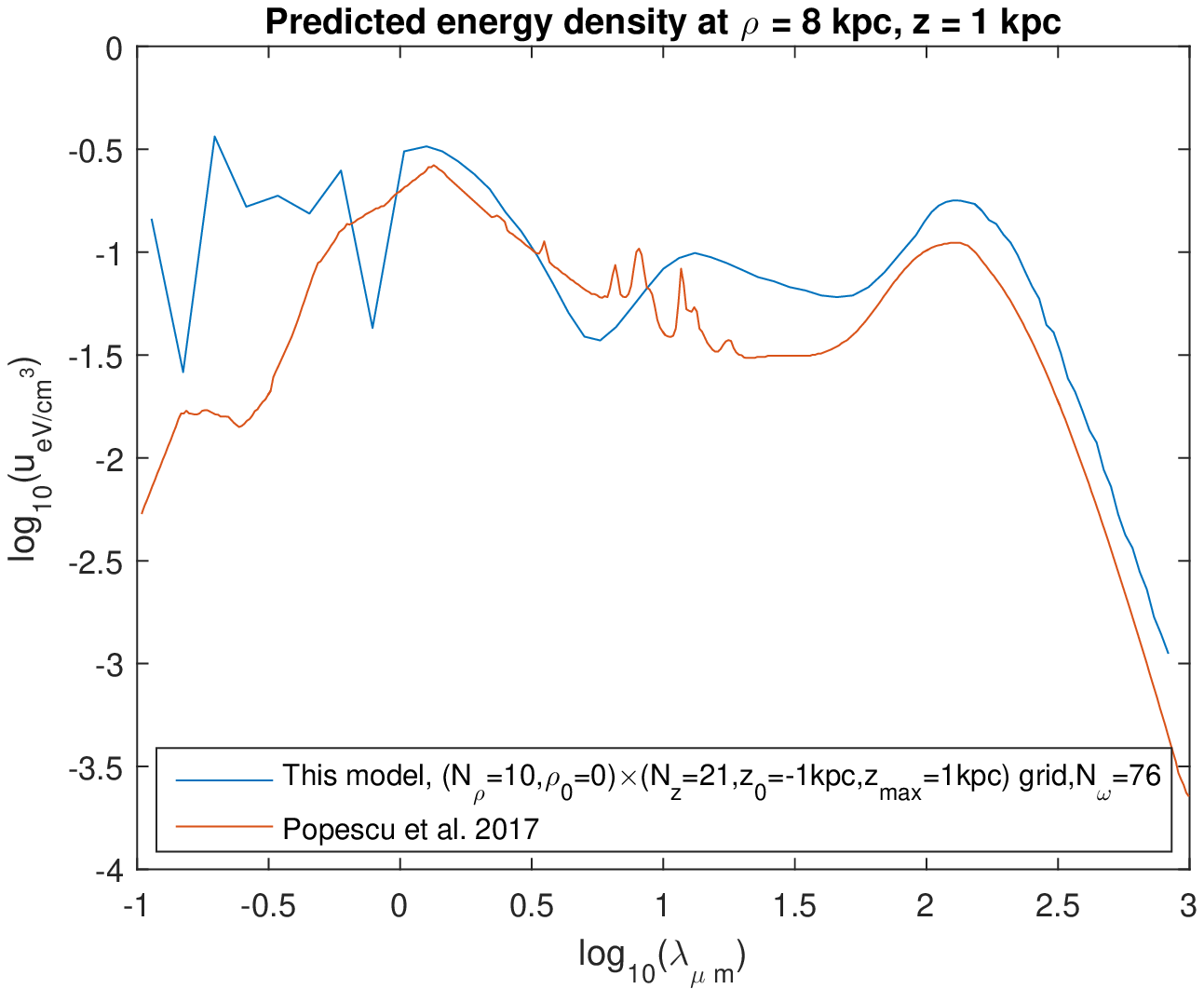}}
\caption{SED at $\rho =8\,$kpc, $z=1\,$kpc, according to this model or a radiation transfer model.}
\label{SED_8_1}
\end{figure}

\begin{figure}[ht]
\centerline{\includegraphics[height=10cm]{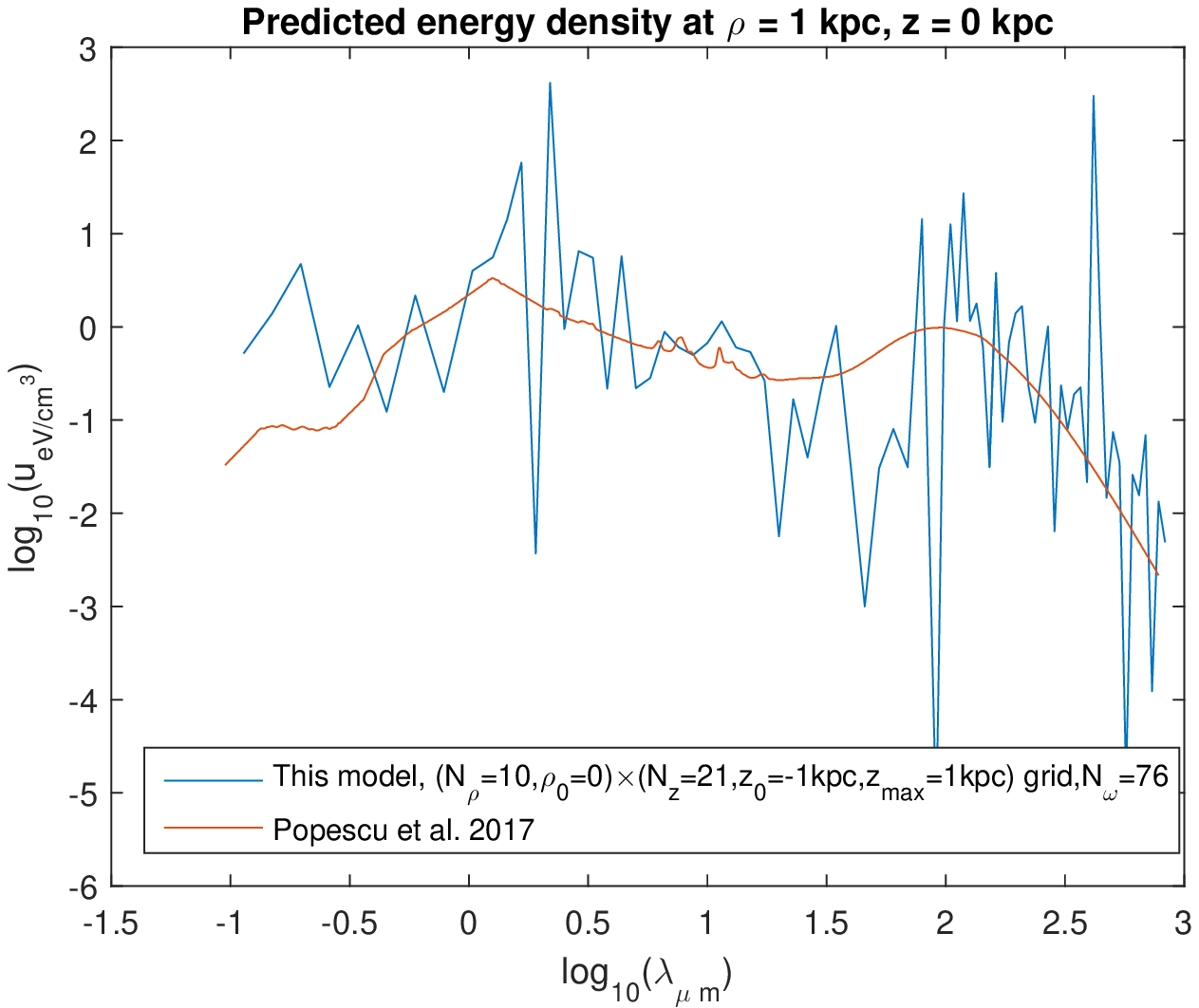}}
\caption{SED at $\rho =1\,$kpc, $z=0\,$kpc, according to this model or a radiation transfer model.}
\label{SED_1_0}
\end{figure}

\begin{figure}[ht]
\centerline{\includegraphics[height=10cm]{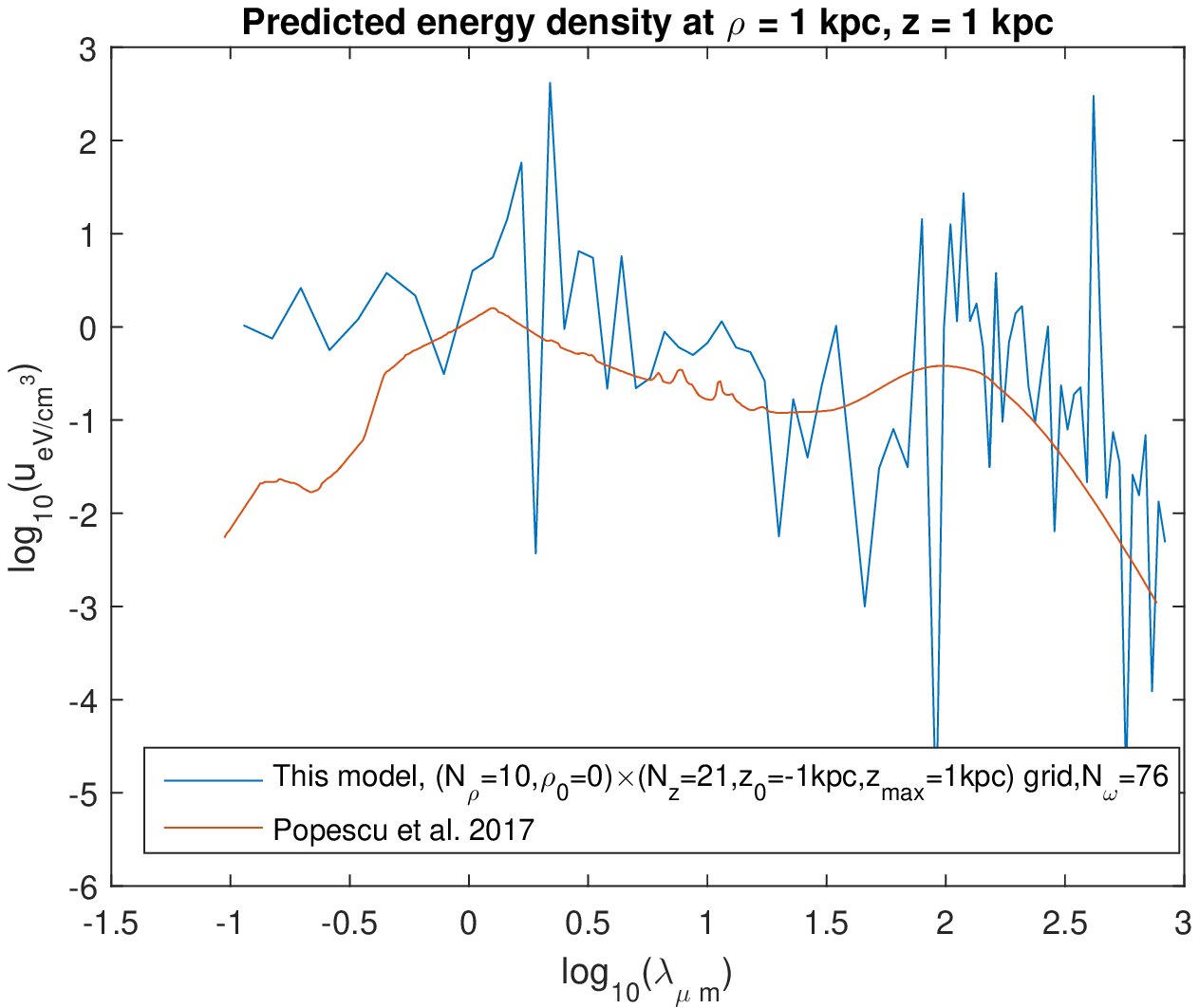}}
\caption{SED at $\rho =1\,$kpc, $z=1\,$kpc, according to this model or a radiation transfer model.}
\label{SED_1_1}
\end{figure}

Then, at any other spatial position ${\bf x}(\rho ,z)$ (in the axially symmetric case considered), the model gives a prediction for the values $u_j({\bf x})$ of the SED corresponding to the frequencies $\omega _j$ (or the wavelengths $\lambda _j$) considered in our calculation. The spatio-temporal grid used was $N_t = 5,\quad N_\rho = 10, \quad N_z = 21$ (thus 1050 events $(t,\rho ,z)$), with $t$ varying from $0$ to $T_0=\lambda _0/c$ with $\lambda _0=10\,\mu \mathrm{m}$, $\rho$ varying from $0$ to $10\,$kpc, and $z$ from $-1\,$ to $+1\,$kpc. Other grids have also been tested, giving very similar results. Moreover, unless otherwise mentioned, the discretization number $N$ was $N=48$. (This is the number of subintervals in which any of the integration intervals $[-K_j,+K_j]$ is subdivided, see Eqs. (\ref{k_n})--(\ref{S_nj}).) Figures \ref{SED_8_0} to \ref{SED_1_1} show the comparison of the SEDs, obtained at four different places in our Galaxy, according to either the thus-adjusted present model, or the radiation transfer model of Popescu {\it et al.} \cite{Popescu-et-al2017}. (Again, the relevant curves, Fig. 9 in Ref. \cite{Popescu-et-al2017}, have been digitalized using CurveUnscan.) \\

Clearly, the general agreement between these two calculations is far from perfect. In particular, our calculation oscillates markedly as function of $\lambda $: at all wavelengths, for the two positions that are closer to the galactic centre ($\rho = 1\,$kpc, Figs. \ref{SED_1_0} and \ref{SED_1_1}); and merely at short wavelengths, for the two positions that are farther from the galactic centre ($\rho = 8\,$kpc, Figs. \ref{SED_8_0} and \ref{SED_8_1}). Nevertheless, this agreement is actually surprisingly good if one remembers how different are the two models. The SED predicted by our model is higher than the one predicted by the radiation transfer model of Ref. \cite{Popescu-et-al2017} at the short wavelengths ($\lambda \lesssim 0.5 \mu $m, or $\log_{10} \lambda \lesssim -0.3 $) --- except for the position ($\rho =8\,$kpc, $z=0\,$kpc), that is, relatively speaking, very close to our local position ($\rho_\mathrm{loc} = 8\,$kpc, $z_\mathrm{loc} = 0.02\,$kpc), where the adjustment is done. The higher levels found with our model at the short wavelengths might have a relation with the fact that the absorption by dust, not taken into account by our model, is strong for these radiations. However, this relation is not obvious, since our model is adjusted on the measured SED at our local position, and that local SED is of course affected by the dust absorption to which the radiations directed towards here have been subjected along their paths in the Galaxy. Actually, one notes also that, in the long wavelength domain as well as in the very short wavelength domain, the SED predicted by this model is, in average, higher than that predicted by the radiation transfer model, at the two positions which have a higher altitude ($z=1\,$kpc): Figs. \ref{SED_8_1} and \ref{SED_1_1}. Comparing Figs. \ref{SED_8_0} and \ref{SED_8_1}, and comparing Figs. \ref{SED_1_0} and \ref{SED_1_1}, it appears that the present model with its current numerical settings predicts a slower decrease of the energy density as $z$ (or rather $\abs{z}$) increases than does the radiation transfer model. Of course the two models are very different, also in their modelling of the Galaxy: a discrete star distribution (for the present model) vs. an analytical, continuous variation of the stellar emissivity as function of $\rho $ and $z$ (for the radiation transfer model of Popescu {\it et al.} \cite{Popescu-et-al2017}). However, the $\rho $ and $z$ distribution of the stars in our model has been built to correspond (at least approximately) with that for the Milky Way \cite{A61}, as has been built the $\rho $ and $z$ dependence of the stellar emissivity function for the radiation transfer model \cite{Popescu-et-al2017}. Of course also, the SEDs of a radiation transfer model (like that of Ref. \cite{Popescu-et-al2017}) are a prediction, too, not a measurement. Nevertheless, a further investigation is necessary to quantify the slower decrease of the SED with increasing $z$ for the present model and to understand the reason for it. \\

\begin{figure}[ht]
\centerline{\includegraphics[height=10cm]{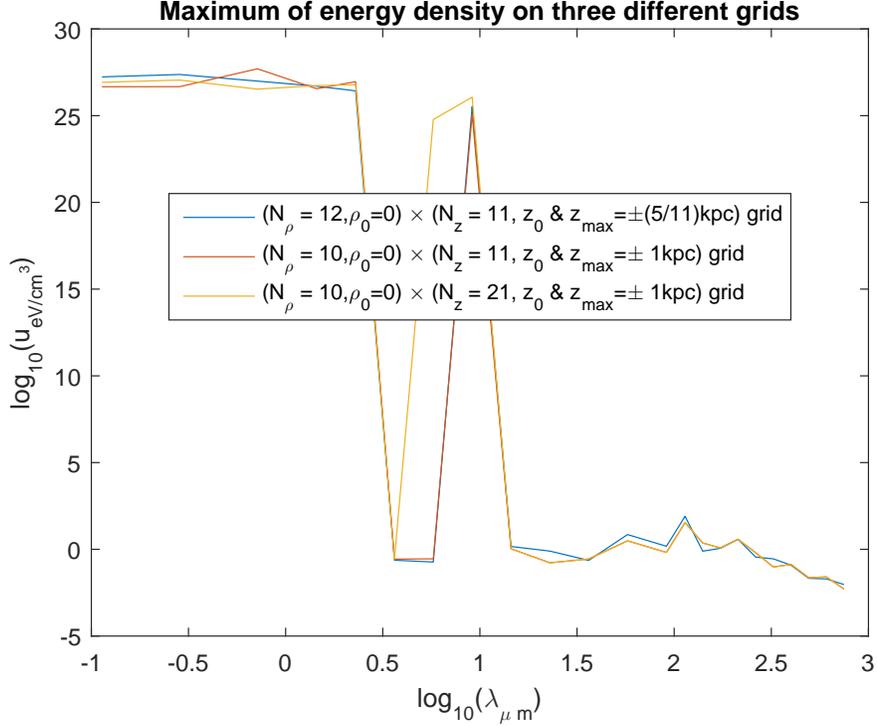}}
\caption{Predicted maximum values of the SED: $N_\omega =23$, three different grids.}
\label{SED_Max}
\end{figure}

\begin{figure}[ht]
\centerline{\includegraphics[height=10cm]{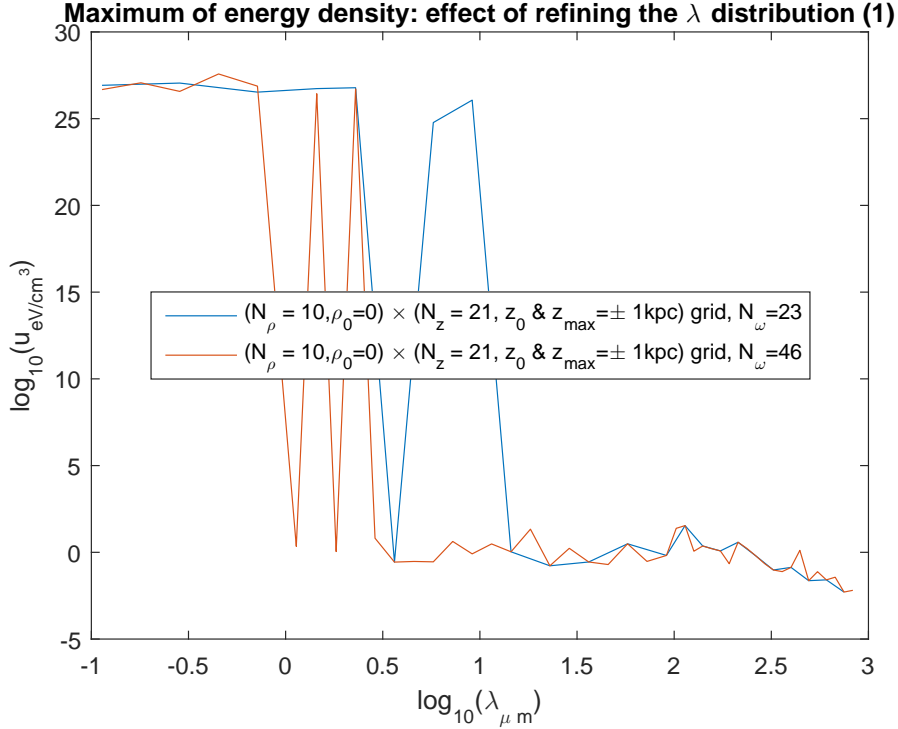}}
\caption{Predicted maximum values of the SED: $N_\omega =23$ vs $N_\omega =46$.}
\label{SED_Max23-46}
\end{figure}

\begin{figure}[ht]
\centerline{\includegraphics[height=10cm]{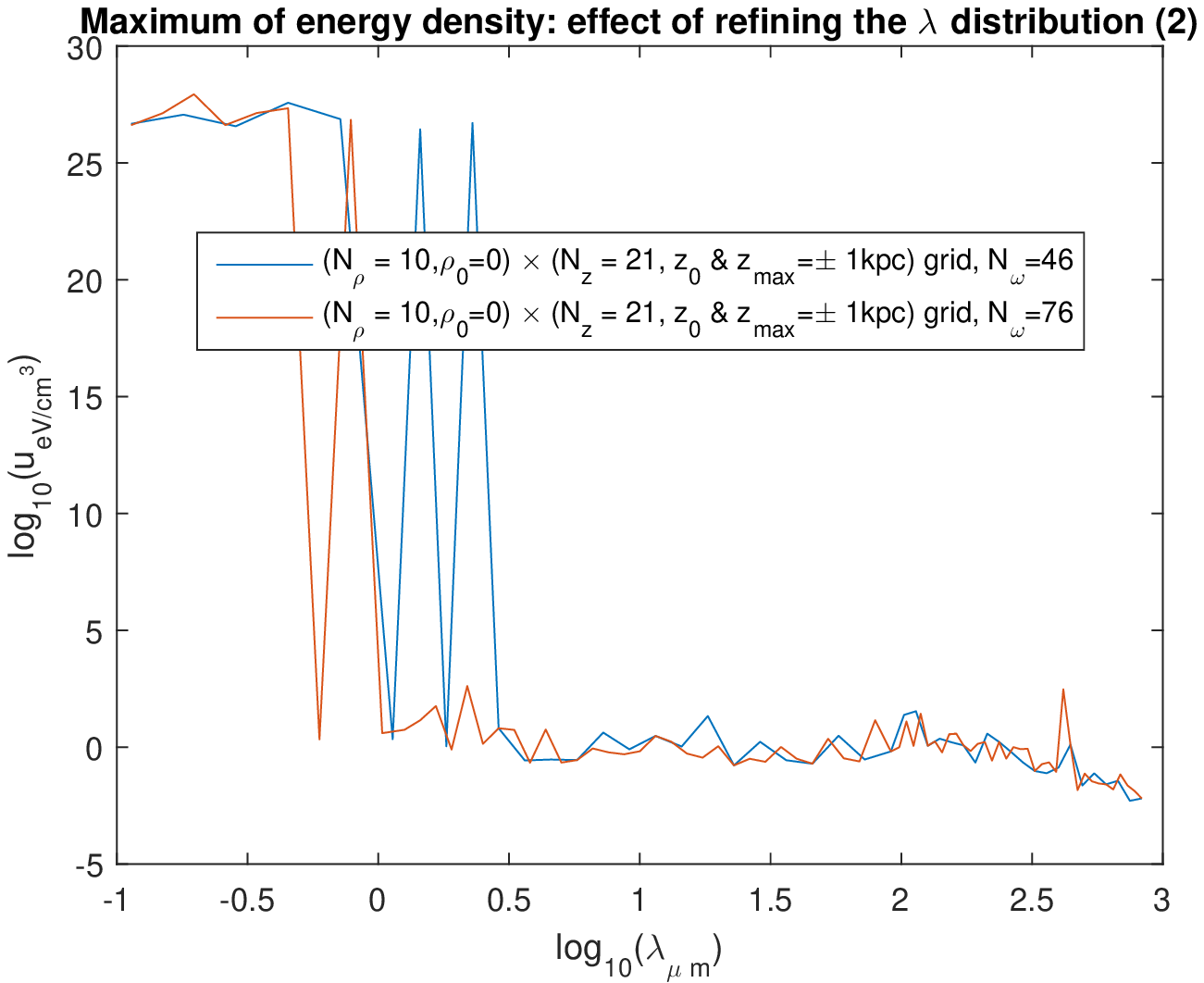}}
\caption{Predicted maximum values of the SED: $N_\omega =46$ vs $N_\omega =76$.}
\label{SED_Max46-76}
\end{figure}

\begin{figure}[ht]
\centerline{\includegraphics[height=10cm]{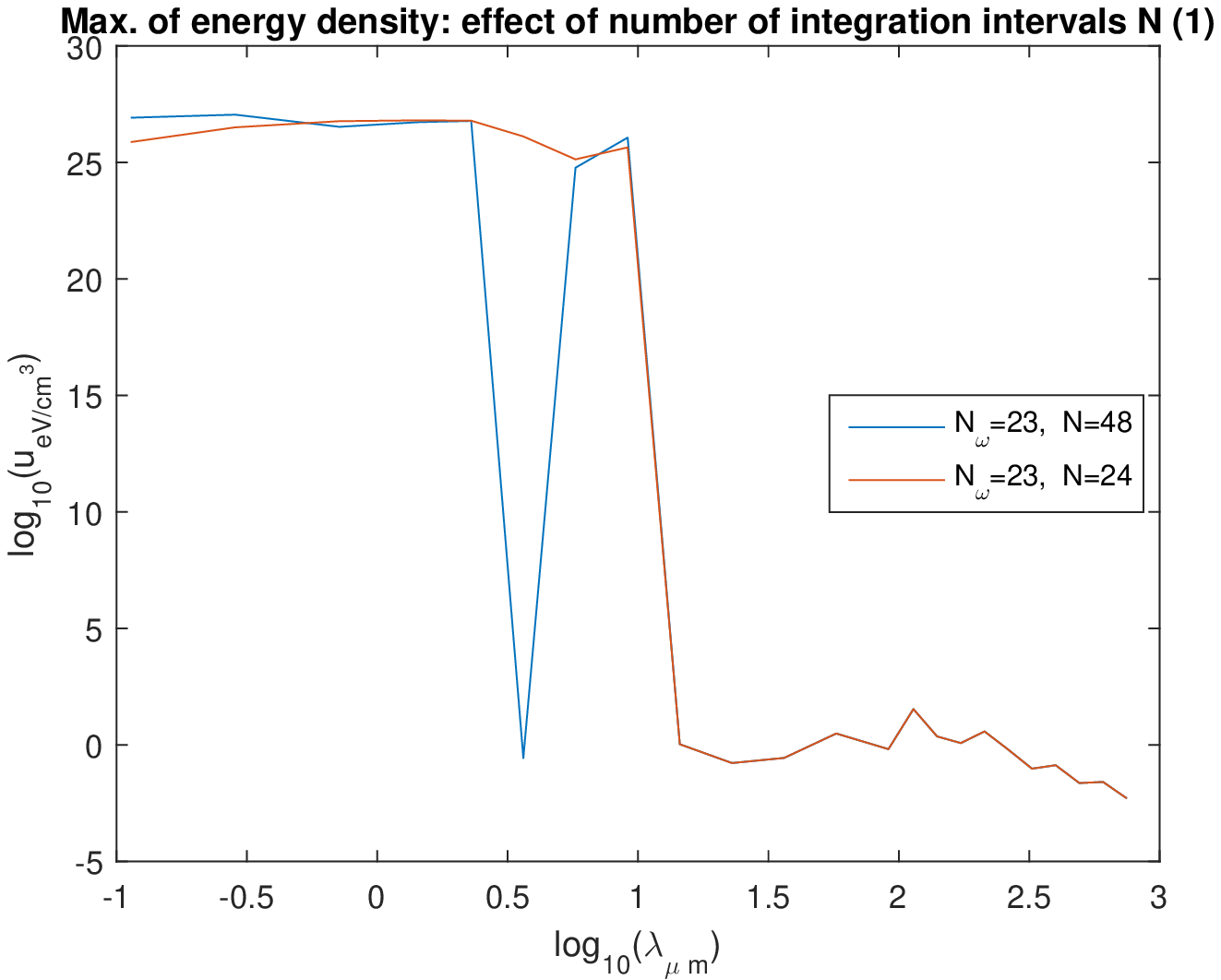}}
\caption{Predicted maximum values of the SED for $N_\omega=23$: $N =48$ vs $N=24$.}
\label{SED_Max_48vs24_23}
\end{figure}

\begin{figure}[ht]
\centerline{\includegraphics[height=10cm]{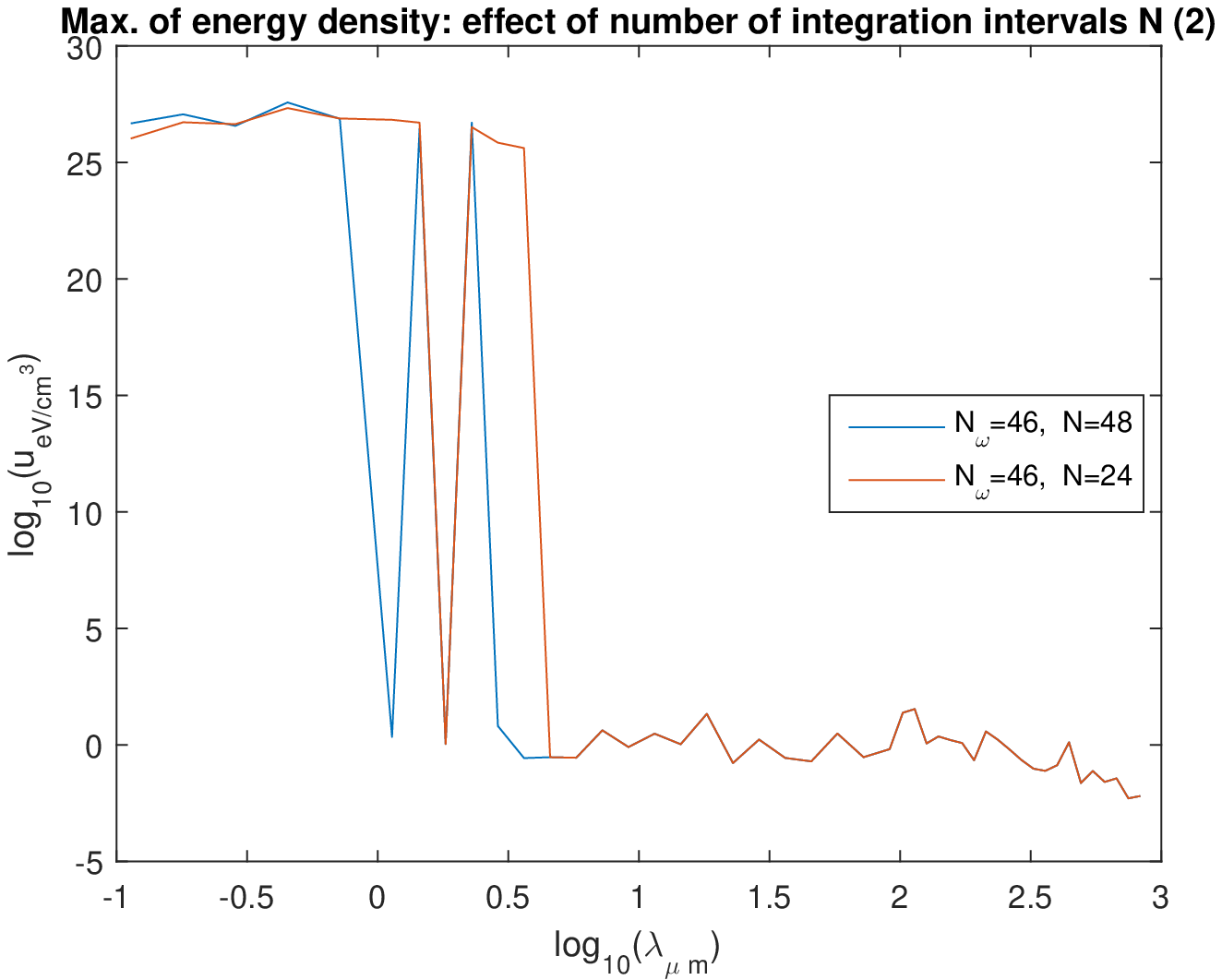}}
\caption{Predicted maximum values of the SED for $N_\omega=46$: $N =48$ vs $N=24$.}
\label{SED_Max_48vs24_46}
\end{figure}

\begin{figure}[ht]
\centerline{\includegraphics[height=10cm]{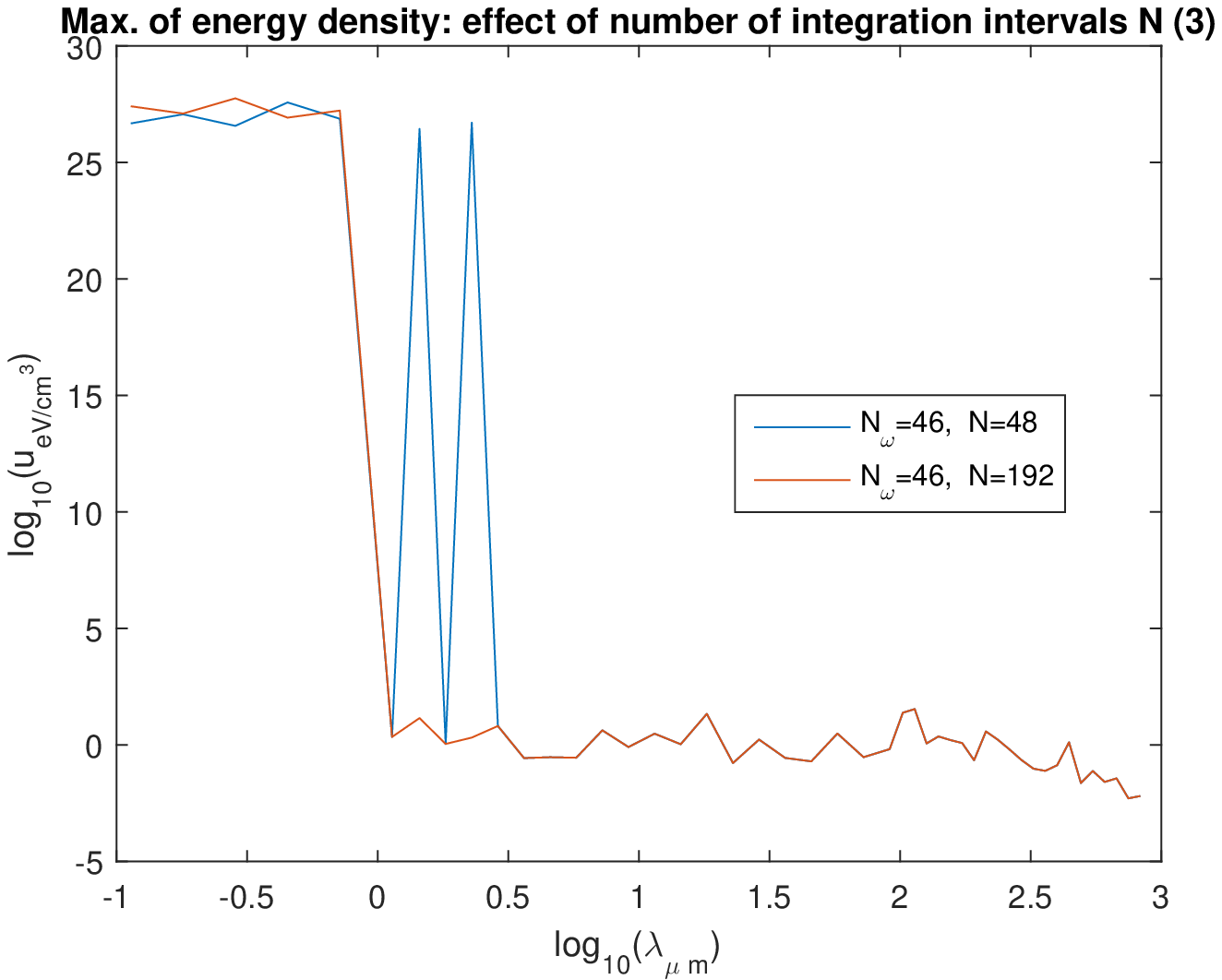}}
\caption{Predicted maximum values of the SED for $N_\omega=46$: $N =48$ vs $N=192$.}
\label{SED_Max_46_48vs192}
\end{figure}

A surprising prediction of our model is that for the maximum values $u_{j\, \mathrm{max}}$ of the SED in our model of the Galaxy --- the SED being calculated at each point on a regular $(\rho ,z)$ grid, as described above. Figure \ref{SED_Max} shows these maximum values for three different spatial grids, with the same set of $23$ wavelengths or frequencies. It is seen that the $u_{j\, \mathrm{max}}$ values are quite similar for the different grids, and these values are extremely high at the short wavelengths, i.e., at high energies of the radiation field. They are found on the $\rho =0$ axis, thus on the central axis of the galaxy. (Actually, the values $u_j(\rho =0,z)$ are fairly independent of the value of $z$ in the domain investigated, i.e., $-1\,\mathrm{kpc} \le z \le 1\,\mathrm{kpc}$.) However, if one refines the frequency mesh, going from $N_\omega =23$ to $N_\omega =46$, and then to $N_\omega =76$, one observes that the very high values of $u_{j\, \mathrm{max}}$ are obtained only at shorter and shorter wavelengths (Figs. \ref{SED_Max23-46} and \ref{SED_Max46-76}). A similar phenomenon is observed when one refines the discretization parameter $N$: for a given frequency mesh, increasing $N$ reduces the domain of the high values of $u_{j\, \mathrm{max}}$ towards smaller values of $\lambda _j$, or, equivalently, decreasing $N$ expands that domain towards larger values of $\lambda _j$ (Figs. \ref{SED_Max_48vs24_23}--\ref{SED_Max_46_48vs192}). The likely common reason for the two effects is that, when one refines either the frequency mesh ($N_\omega $ increased) or the discretization of the integration intervals ($N$ increased), then the number of the solved-for parameters $S_{nj} \ (n=0,...,N;\, j=1,...,N_\omega )$ increases. In most of the settings investigated in the present work, this number: $N_\mathrm{para}= (N+1) \times N_\omega $, is actually larger than the number $N_\mathrm{data}$ of ``data", that is the number of points in the spatio-temporal grid,
\footnote{\
Indeed, at each of those points one calculates the function on the l.h.s. of Eq. (\ref{Psi'}). These are the data which have to be fitted by using the function on the r.h.s., that depends on the solved-for parameters $S_{nj}$.  
}
thus $N_\mathrm{data}=N_t\times N_\rho \times N_z$. We have $N_\mathrm{data}=1050$ for the grid used in this work, except for the two grids used for the first two curves in Fig. \ref{SED_Max}, which give respectively $N_\mathrm{data}=660$ and $N_\mathrm{data}=550$. (We have $N_t=5$ also for those two grids.) Hence, we are often in a situation of ``overfitting". 
\footnote{\
However, the least-squares solver: the Matlab backslash operator {\sf mldivide}, puts to zero many of the $S_{nj}$ parameters for reason of insufficient numerical accuracy, so that their effective number was $150$ to $200$, thus significantly smaller than $N_\mathrm{data}$.
}
Higher values of the ratio $R= N_\mathrm{para}/N_\mathrm{data}$ correspond with a more overfitted situation, in which one expects the predictions to be less accurate. Now, on each among Figs. \ref{SED_Max23-46}--\ref{SED_Max46-76} and \ref{SED_Max_48vs24_23}--\ref{SED_Max_46_48vs192}, we saw that the domain of the low values of $u_{j\, \mathrm{max}}$ increases when $N_\mathrm{para}$ increases, the  number $N_\mathrm{data}$ being fixed at $N_\mathrm{data}=1050$ for these five figures --- thus we saw that the domain of the low values of $u_{j\, \mathrm{max}}$ increases when $R$ increases, i.e., when more overfitting occurs. Therefore, it is the {\it low} values of $u_{j\, \mathrm{max}}$ which are likely to be a numerical artefact, not the extremely high values.
%Now, higher values of the ratio $R= N_\mathrm{para}/N_\mathrm{data}$ correspond with a more overfitted situation, in which one expects the predictions to be less accurate. According to this tentative explanation, it is the {\it low} values of $u_{j\, \mathrm{max}}$ (whose domain has been seen to increase when the ratio $R$ increases), which are likely to be a numerical artefact, not the extremely high values. 
Note that this effect can be seen also on Fig. \ref{SED_Max}: $N$ and $N_\omega $, hence $N_\mathrm{para}$, are the same for the three curves. So the third and finest grid $(N_\rho =10)\times (N_z=21)$ leads to a lower $R$ than do the other two grids $(N_\rho =12)\times (N_z=11)$ and $(N_\rho =10)\times (N_z=11)$ --- and accordingly the domain of the high values of $u_{j\, \mathrm{max}}$ is larger for the third curve. On the other hand, the effect of overfitting seems less important at a distance from the axis (where the values of the SED are not high at all): e.g., if one changes $N_\omega $ from $76$ to $23$, the SEDs considered in Figs. \ref{SED_8_0}--\ref{SED_1_1} stay similar: they show a smaller number of oscillations due to the smaller number of points, but the amplitude of the oscillations is comparable. Tentatively, we may understand this by finding it natural that the very high numbers are more sensitive to overfitting. In summary, the status of the very high values of the maxima of $u_j(\rho ,z)$: physical prediction of the model or numerical artefact, is not really clear at the present stage of this work --- but the foregoing discussion, with different examples that all show an effect which can be attributed to the ``amount of overfitting", seems to indicate that those high values might be a true prediction of the model.

%%%%%%%%%%%%%%%%%%%%%%%%%%%%%%%%%%%%%%%%%%%%%%%%%%%%%%%%%%%%%%%%%%%%%%%%%%%%%%%%
\section{Conclusion}
%%%%%%%%%%%%%%%%%%%%%%%%%%%%%%%%%%%%%%%%%%%%%%%%%%%%%%%%%%%%%%%%%%%%%%%%%%%%%%%%

In this work, the Maxwell model of the interstellar radiation field \cite{A61} has been first checked for its predictions of the spatial variation of the spectral energy distributions (SEDs) in our Galaxy. The model has been adjusted by asking that the SED predicted at our local position in the Galaxy, Eq. (\ref{Udiscrete}) with $\rho =\rho_\mathrm{loc} := 8$ kpc and $z=z_\mathrm{loc} := 0.02$ kpc, coincide with the observations collected through different spatial missions \cite{Henry-et-al1980,Arendt-et-al1998,Finkbeiner-et-al1999}. Then the predictions of the model for other positions in the Galaxy (for which, of course, we have no real observation) have been compared with the predictions obtained by using a recent radiation transfer model \cite{Popescu-et-al2017}. The two predictions do not differ too much in magnitude, even though the predictions of the present model oscillate rather strongly as function of the wavelength, especially at short wavelengths and not very far (1 kpc) from the axis of the Galaxy. Also, the SED decreases more slowly with increasing altitude for the present model. The values of the SED reach their maximum on the axis itself of the Galaxy; that maximum is extremely high at short wavelengths. These surprisingly high values occur in a larger or smaller domain of wavelengths, depending on the settings of the calculation. However, this dependence is governed by the ``amount of overfitting": less overfitting increases the range of the high values. This makes it plausible that the high values might be a true prediction of the model. The next work will aim at improving the numerical procedure, in order: (i) to try to establish if the high values are indeed a true prediction of the model; (ii) to check if a numerically improved model could lead to oscillations of a smaller amplitude; (iii) to quantify the slower decrease of the SED with increasing altitude for the present model and to find the reason for it. Nevertheless, the present results: (1) show that the prediction, by the Maxwell model of the ISRF, of the spatial variation of the SED in the Galaxy, is quite comparable in magnitude to that provided by a radiation transfer model (except perhaps for high energies when the point is very close to the galaxy's axis); and therefore (2) indicate that an order-of-magnitude estimate of the ``interaction energy" predicted \cite{A57} by an alternative, scalar theory of gravitation, should be indeed possible with the present model in a future work.\\

{\bf Acknowledgement.} I am grateful to the reviewer for his (or her) careful study of my manuscript, and for the important remarks and questions. In particular, the remark on the likely overfitting, and the question about the reason for the dependence of the maximum of energy density on the frequency mesh, taken together, led me to improve significantly the discussion of the maximum values of the SED. Also, the question about the reason why, on Fig. 3, this model's curve shifts upward from the radiation transfer model in the long wavelength regime, led me to comment on the dependence of the SED with the altitude $z$.

%%%%%%%%%%%%%%%%%%%%%%%%%%%%%%%%%%%%%%%%%%%%%%%%%%%%%%%%%%%%%%%%%%%%%%%%%%%%%%%%
\appendix
%%%%%%%%%%%%%%%%%%%%%%%%%%%%%%%%%%%%%%%%%%%%%%%%%%%%%%%%%%%%%%%%%%%%%%%%%%%%%%%%
%%%%%%%%%%%%%%%%%%%%%%%%%%%%%%%%%%%%%%%%%%%%%%%%%%%%%%%%%%%%%%%%%%%%%%%%%%%%%%%%
\section{Appendix: Extension of the explicit representation to a finite spectrum}\label{Extension}
%%%%%%%%%%%%%%%%%%%%%%%%%%%%%%%%%%%%%%%%%%%%%%%%%%%%%%%%%%%%%%%%%%%%%%%%%%%%%%%%

In order to present this extension of the result \cite{A60} (the latter being recalled in Subsect. \ref{Explicit}), it is necessary to write this result in terms of mappings. Consider the vector space $\mathcal{S}^\omega $ (respectively $\mathcal{F}^\omega $) made of those solutions $\Psi $ of the scalar wave equation (respectively, made of those source-free Maxwell fields  $\Mat{F}=({\bf E},{\bf B})$) that are axisymmetric and time-harmonic with frequency $\omega $. 
\footnote{\
As the domain of definition of the fields in $\mathcal{S}^\omega $ and $\mathcal{F}^\omega $, we take the whole spacetime. More exactly, since we assume axial symmetry, we assume that they are defined for the whole range of values of the spacetime variables $t,\rho,z$, thus in the domain $\Omega $: ($t, z \in \mathbb{R}$, $\rho \in \mathbb{R}_+$).
}
We define two mappings $Z_1$ and $Z_2$ from $\mathcal{S}^\omega $ into $\mathcal{F}^\omega $: given $\Psi $ in $\mathcal{S}^\omega $, $Z_1(\Psi )$ is the EM field given by Eqs. (\ref{Bfi})--(\ref{Ez}); and $Z_2(\Psi )$ is the EM field given by Eqs. (\ref{Bfi})--(\ref{Ez}), followed by Eq. (\ref{dual}). Obviously, $Z_1$ and $Z_2$ are linear. We can see that $Z_1(\Psi )$ belongs to $\mathcal{F}_1^\omega $, the subspace of $\mathcal{F}^\omega $ with $E_\phi =B_\rho =B_z=0$, whereas $Z_2(\Psi )$ belongs to $\mathcal{F}_2^\omega $, the subspace of $\mathcal{F}^\omega $ with $B_\phi =E_\rho =E_z=0$. The {\it uniqueness} of the decomposition (\ref{Decompos}) is just that of the EM fields $Z_1(\Psi ) = \Mat{F}_1 \in \mathcal{F}_1^\omega$ and $Z_2(\Psi )=\Mat{F}_2 \in \mathcal{F}_2^\omega$ (if they exist), such that the starting EM field $\Mat{F} \in \mathcal{F}^\omega$ is $\Mat{F}=\Mat{F}_1+\Mat{F}_2$. This results immediately from the complementarity of the non-zero components, noted in Subsect. \ref{Explicit}, and inherent in the definition of $\mathcal{F}_1^\omega$ and $\mathcal{F}_2^\omega$: the non-zero components of $\Mat{F}_1 $ and $\Mat{F}_2$ must be just the corresponding components of $\Mat{F}$, hence are known from the data of $\Mat{F}$. Henceforth, for {\it any} EM field $\Mat{F}$, we will note $\Mat{F}_1$ (respectively $\Mat{F}_2$) the field whose components $E_\phi, B_\rho, B_z$ are zero (respectively the field whose components $B_\phi, E_\rho,E_z$ are zero), the other components being those of $\Mat{F}$. In contrast with the uniqueness, the {\it existence} of the decomposition (\ref{Decompos}) is a rather strong result, which amounts to state that the mapping $Z$ from $\mathcal{S}^\omega \times \mathcal{S}^\omega $ to $\mathcal{F}^\omega $ defined by
\be
Z(\Psi ,\Psi ')= Z_1(\Psi )+Z_2(\Psi') 
\ee
is surjective. This also amounts to state that, given any EM field $\Mat{F} \in \mathcal{F}^\omega $, there exist $\Psi $ and $\Psi '$ in $\mathcal{S}^\omega $, such that $\Mat{F}_1=Z_1(\Psi )$ and $\Mat{F}_2=Z_2(\Psi' )$. \\

The general axisymmetric solutions of either the scalar wave equation or the source-free Maxwell equations are got by summing time-harmonic axisymmetric solutions of the respective equations. %We consider only the totally propagating situation, since it is the relevant one for the ISRF. 
We restrict ourselves to such solutions having a {\it finite set of frequencies,} say $(\omega _j)\ (j=1,...,N_\omega )$, for simplicity \cite{A61,A60}. The axisymmetric solutions of the wave equation with this set of frequencies form the vector space $\mathcal{S}:=\mathcal{S}^{\omega_1}\oplus ...\oplus \mathcal{S}^{\omega_{N_\omega }}$. The axisymmetric solutions of the source-free Maxwell equations with this set of frequencies form the vector space $\mathcal{F}:=\mathcal{F}^{\omega_1}\oplus ...\oplus \mathcal{F}^{\omega_{N_\omega }}$. The sums are direct because the decomposition in a finite sum of time-harmonic fields is of course unique. Therefore, the linear mappings $Z_1$ and $Z_2$ extend immediately to ones from $\mathcal{S}$ to $\mathcal{F}_1:=\mathcal{F}_1^{\omega_1}\oplus ...\oplus \mathcal{F}_1^{\omega_{N_\omega }}$ or respectively $\mathcal{F}_2:=\mathcal{F}_2^{\omega_1}\oplus ...\oplus \mathcal{F}_2^{\omega_{N_\omega }}$, by linearity, and so does the theorem. That is: for any EM field $\Mat{F}=\Mat{F}^{(1)}+...+\Mat{F}^{(N_\omega) }$ in the direct sum $\mathcal{F}$, the result valid for the time-harmonic case ensures that there exist $2N_\omega $ scalar potentials: $\Psi ^{(1)},\Psi'^{(1)}\in \mathcal{S}^{\omega_1},...,\Psi^{(N_\omega) },\Psi'^{(N_\omega) }\in \mathcal{S}^{\omega_{N_\omega }}$, such that 
\be
\Mat{F}_1^{(1)}=Z_1(\Psi ^{(1)}),\Mat{F}_2^{(1)}=Z_2(\Psi'^{(1)}),...,\Mat{F}_1^{(N_\omega) }=Z_1(\Psi^{(N_\omega) }), \Mat{F}_2^{(N_\omega) }=Z_2(\Psi'^{(N_\omega) }).
\ee
Hence, for any EM field $\Mat{F} \in \mathcal{F}$, there are indeed two scalar potentials $\Psi =\Psi^{(1)}+...+\Psi^{(N_\omega) }$ and $\Psi' =\Psi'^{(1)}+...+\Psi'^{(N_\omega) }$ in the direct sum $\mathcal{S}$, such that 
\be\label{F_1,F_2}
\Mat{F}_1 = Z_1(\Psi ):= Z_1(\Psi ^{(1)})+...+Z_1(\Psi^{(N_\omega) }), \quad \Mat{F}_2 = Z_2(\Psi' ):= Z_2(\Psi'^{(1)})+...+Z_2(\Psi'^{(N_\omega) }),
\ee
and thus, finally,
\be\label{F_1+F_2}
\Mat{F}= \Mat{F}_1+\Mat{F}_2 = Z_1(\Psi )+Z_2(\Psi' ) :=Z(\Psi ,\Psi ').
\ee
In the totally propagating case, we have $\Psi^{(j)} = \psi _{\omega_j\ S_j}$ and $\Psi'^{(j)}= \psi _{\omega_j\ S'_j}$, for some $2N_\omega $ functions $S_j(k), S'_j(k), k\in[-K_j,+K_j] \ (j=1,...,N_\omega)$, with $K_j:=\frac{\omega_j }{c}$.

%%%%%%%%%%%%%%%%%%%%%%%%%%%%%%%%%%%%%%%%%%%%%%%%%%%%%%%%%%%%%%%%%%%%%%%%%%%%%%%%
\end{document}